\documentstyle[12pt]{article}
\def\kpc{\,{\rm kpc}}

\def\cmm2{{\,\rm cm^{-2}}}
\def\cm2{{\,{\rm cm}^2}}
\def\cmm3{{\,{\rm cm}^{-3}}}
\def\gcmm3{{\,{\rm g\,cm^{-3}}}}
\def\kms{\,{\rm km\,s^{-1}}}

\def\fun#1#2{\lower3.6pt\vbox{\baselineskip0pt\lineskip.9pt
  \ialign{$\mathsurround=0pt#1\hfil##\hfil$\crcr#2\crcr\sim\crcr}}}
\begin{document}
\pagestyle{empty}
\begin{center}
\bigskip
\rightline{FERMILAB--Pub--94/381-A}
\rightline{astro-ph/9411073}
\rightline{submitted to {\it Physical Review Letters}}

\vspace{.2in}
{\Large \bf Microlensing and Halo Cold Dark Matter}
\bigskip

\vspace{.2in}
Evalyn I.~Gates,$^{1,2}$ Geza Gyuk,$^{2,3}$ and Michael S. Turner$^{1,2,3}$\\
\vspace{.2in}
{\it $^1$Department of Astronomy \& Astrophysics\\
Enrico Fermi Institute, The University of Chicago, Chicago, IL~~60637-1433}\\

\vspace{0.1in}
{\it $^2$NASA/Fermilab Astrophysics Center\\
Fermi National Accelerator Laboratory, Batavia, IL~~60510-0500}\\

\vspace{0.1in}
{\it $^3$Department of Physics, Enrico Fermi Institute\\
The University of Chicago, Chicago, IL~~60637-1433}\\

\end{center}

\vspace{.3in}

\centerline{\bf ABSTRACT}
\bigskip

We discuss the implications of the more than 50
microlensing events seen by the EROS, MACHO, and OGLE collaborations
for the composition of the halo of our galaxy.  The event rates indicate
that the halo mass fraction in MACHO's is less than 30\%,
consistent with expectations for a universe whose
primary component is cold dark matter.  We caution that
the uncertainties are such that a larger MACHO fraction
cannot yet be excluded.

\newpage
\pagestyle{plain}
\setcounter{page}{1}
\newpage

In 1986 Paczynski suggested microlensing as a probe of dark (or very faint)
stars in our galaxy \cite{pac} (referred to generically
as MAssive Compact Halo Objects, or MACHOs).  In the past year
more than 50 microlensing events
have been reported.  (For microlensing the two images are too
close to be resolved; instead, the combined light leads to
an achromatic, time-symmetric brightening.)  The EROS collaboration
has seen two events in the direction of the Large Magellanic Cloud (LMC)
\cite{eros}; the OGLE collaboration has seen 12
events in the direction of the galactic bulge \cite{ogle}; and
the MACHO collaboration has seen three events in the direction
of the LMC and more than 40 in the direction of the galactic bulge
\cite{macho}.

The study of microlensing toward the galactic bulge
mainly probes the structure of the inner galaxy,
while microlensing toward the LMC mainly probes the dark halo
\cite{pac,griest}.  The probability that a given star is being
microlensed by a foreground object is referred to as the optical depth
for microlensing ($\equiv \tau$).  Expectations for the
bulge were $\tau_{\rm BULGE}\simeq 1\times 10^{-6}$,
largely due to lower-main-sequence
stars in the disk \cite{bulpred}; OGLE reports an optical
depth that is about a factor of three larger, $\tau_{\rm OGLE} =
3.3\pm 1.2 \times 10^{-6}$ \cite{ogle}, and the rate observed
by MACHO may be even higher \cite{machobulge}.  Expectations for an all-MACHO
halo were $\tau_{\rm LMC}\simeq 5\times 10^{-7}$ \cite{griest}
(because data poorly constrain
the halo, uncertainties in this estimate are large,
almost a factor two either way \cite{gates,griestnew,megomodels}).
Based upon 9 million-star years of observations, the three events
observed, and estimates of their efficiencies (between 20\% and 40\%)
\cite{machoproc}, the MACHO data indicate that $\tau_{\rm MACHO}
\simeq 1\times 10^{-7}$.  The EROS data indicate a similar optical
depth \cite{eros}.

Evidence that spiral galaxies, including our own, are embedded in
extended, dark (roughly spherical) halos comes from galactic
rotation curves, the study of satellite galaxies and binary galaxies,
the kinematics of the globular clusters in our galaxy,
the warping of galactic disks, and the flaring of neutral
hydrogen gas associated with disks \cite{Zarits,binney,frenk}.
Galactic halos are repositories for both nonbaryonic dark matter
(mainly slowly moving particles, or cold dark matter, since fast moving
particles such as light neutrinos move too fast to be captured) and
baryonic dark matter.  Experimental efforts to detect
nonbaryonic dark matter have focussed on our own halo \cite{expt}.
Determining the mass fraction
of the halo in baryons is crucial for estimating the amount of
nonbaryonic matter that may exist in our galaxy.

Our purpose here is to use the microlensing data to draw conclusions about
the MACHO fraction of the halo---and from it the fraction of the halo
that could be particle dark matter.   Since the expected microlensing rate
in the direction of the LMC depends upon galactic modelling
\cite{gates,griestnew,megomodels} and the LMC microlensing statistics are
small, we adopt the following strategy.  The rotation curve,
local projected mass density, distribution
of luminous material in the disk and bulge, and bulge microlensing rate
are used to constrain the halo model.  For acceptable
galactic models we calculate $\tau_{\rm LMC}$, and then,
based upon the observed LMC rate, make inferences about the
MACHO fraction of the halo, concluding that it must less than about 30\%.

Models of the Milky Way have three major components \cite{glxmodels},
a central bulge, a disk, and a spherical halo, with large
uncertainties in the parameters that define all
three.  The basic picture of the bulge has evolved
from spherical to recent indications that it
may be closer to a bar \cite{manybars}.  We follow
Dwek et al. \cite{dwek} who have utilized DIRBE surface brightness
observations to construct a triaxial bulge model:
\begin{equation}
\rho_{\rm BAR} =  {M_0 \over 8 \pi abc} e^{-s^2/2},  \qquad
s^4 = \left[ {x^2\over a^2} + {y^2\over b^2}  \right] ^2 + {z^4 \over c^4} ,
\end{equation}
where the bulge mass $M_{\rm BAR} = 0.82 M_0$, the scale lengths
$a = 1.49\kpc$, $b = 0.58\kpc$ and $c = 0.40\kpc$, and the long axis
is oriented at an angle of about $10^{\circ}$ with respect to
the line of sight toward the galactic center.   The bulge
mass is not well determined, and we consider $M_{\rm BAR}=1,$ 2, 3, and
$4\times 10^{10} M_\odot$ \cite{glxmodels,barmass}.

The bulk of the luminous matter in the disk follows a double exponential
distribution \cite{GWK}.  There is some evidence that the disk
has both a ``thick'' and a ``thin'' component \cite{GWK}.
We take the sum of a ``fixed,'' thin luminous disk,
\begin{equation}
\rho_{\rm LUM}(r,z) = {\Sigma_{\rm LUM}\over 2h}\,
\exp [-(r-r_0)/r_d] e^{-|z|/h},
\end{equation}
with scale length $r_d \sim 3.5\kpc$, scale height
$h = 0.3\kpc$, and local projected mass density
$\Sigma_{\rm LUM} = 25 M_\odot\,{\rm pc}^{-2}$ \cite{lumdisk},
a ``variable'' component.  For the variable component
we use the same scale length, but vary the thickness,
very thin ($h=0.15\kpc$), thin ($h=0.3\kpc$) and thick ($h=1.5\kpc$), and
local projected mass density $\Sigma_{\rm VAR}$.
We also explore a variable component whose surface density scales as $1/r$
rather than exponentially \cite{1/r}.  A number of studies of
stars motions constrain the local projected mass density within a
distance of $0.3\kpc - 1.1\kpc$ of the galactic plane \cite{sigma0}.
As a conservative bound we require that $\Sigma_{\rm TOT}(1\kpc )
= \int_{-1\kpc}^{1\kpc}\rho (r_0, z)dz = 25 - 100 M_\odot\,{\rm pc}^{-2}$;
this implies the variable component of the disk must satisfy
$\Sigma_{\rm VAR}(1\kpc ) \le 75M_\odot\,{\rm pc}^{-2}$.

The third component is the halo.
We assume independent isothermal distributions for
the MACHOs and cold dark matter with core radii $a_i = 2,$ 4, 8, and $16\kpc$,
\begin{equation}
\rho_{\rm HALO,i} = {a_i^2+r_0^2 \over
a_i^2 +r^2}\, \rho_{0,i}\ ,
\end{equation}
where $i=$ MACHO, cold dark matter and $\rho_{0,i}$ is
the local mass density of component $i$.
More complex halo models are possible; e.g.,
flattened halos \cite{griestnew,megomodels}.  We do not expect such
refinements to significantly alter our basic conclusions;
they only serve to increase slightly the theoretical uncertainties.

The average optical depth for microlensing a distant
star by a foreground star is \cite{griest}
\begin{equation}
\tau = {4 \pi G\over c^2} {\int^\infty_0 ds \rho (s)
\int^s_0 dx \rho (x) {x(s-x)/ s} \over \int^\infty_0 ds \rho (s)},
\end{equation}
where $\rho$ is the mass density in stars,
$s$ is the distance to the star being lensed, and $x$ is the
distance to the lens \cite{lenses}.  In estimating the optical depth toward the
bulge, we consider lensing of bulge stars by both disk and bulge objects;
for the LMC we consider lensing of LMC stars by halo and disk objects.
We have assumed that the threshold for the detection
of microlensing is a brightening of 1.34 (which roughly
corresponds to the experimental thresholds).  Further, while
the microlensing rate more closely describes what is measured,
it depends upon detailed knowledge of the velocity distribution
of the lenses, and previous analyses \cite{gates,griestnew}
have found that optical depth correlates well with lensing rate.

Kinematic constraints to the galactic model come from the circular rotation
speed at our position ($\equiv v_c$) and the requirement that the
rotation curve be approximately flat between about $4\kpc$ and $18\kpc$.
We adopt the IAU value of $220\kms$ for $v_c$ with
an uncertainty of $\pm 20\kms$, and we take our distance
from the galactic center to be $r_0=8.0,$ 8.5, and $9.0\kpc$.
For the flatness constraint we follow our previous work \cite{gates}
in requiring that the total variation in $v(r)$ be less than $14\% $
over the aforementioned range.

We construct our suite of viable models as follows.
Starting with a disk (exponential or $1/r$) with local surface
density $\Sigma_{\rm VAR}$ and a bulge of mass $M_{\rm BAR}$
we compute $\tau_{\rm BULGE}$, the optical depth to Baade's window,
galactic coordinates ($1^{\circ} ,-4^{\circ}$),
and the contributions of the disk and bulge to the
rotation curve at $r=r_0$.  For a choice of halo parameters
this then determines the local halo density, the full rotation curve,
and the optical depth to the LMC.  We deem a model viable
if:  $(a)$ $\tau_{\rm BULGE} \geq 2.0\times 10^{-6}$, $(b)$ the
rotation curve is sufficiently flat, and $(c)$ $\tau_{\rm LMC}$ is
in the range $0.5 - 2.0\times 10^{-7}$.
The last condition primarily constrains the baryonic
mass fraction and does not eliminate many models.

Our results are summarized in Figs.~1-4.  We find that in general
the disk alone does not provide sufficient lensing to explain
the event rate seen toward the bulge.
While $\tau_{\rm BULGE}$ increases with $\Sigma_{\rm VAR}$,
$\Sigma_{\rm VAR}$ reaches its upper bound of $75M_\odot\,
{\rm pc}^{-2}$ or the rotation-curve constraint is violated
before $\tau_{\rm BULGE} = 2\times 10^{-6}$.  For
fixed $\Sigma_{\rm VAR}$ the disk density along the line of sight
is about the same, and the difference between a thick and thin disk
is negligible.  However, there
is less mass in a thin disk and a larger $\Sigma_{\rm VAR}$ is
permitted without violating the rotation-curve constraint.
A very thin disk ($h=0.15\kpc$) does not help further:
the line of sight to Baade's window
passes above most of the disk material.
The results for disks with a $1/r$ density distribution are not
significantly different from those of exponential disks.

The bar is a much more efficient source of
lensing \cite{barlens}.  In all viable models a bar mass of at least
$2\times 10^{10} M_\odot$ is required; if $\tau_{\rm BULGE}$
is determined to be greater than $3\times 10^{-6}$,
as may well be the case, a bar mass of at least $3\times 10^{10} M_\odot$
seems unavoidable.  Should the optical depth be $4\times 10^{-6}$,
a bar mass of $4\times 10^{10} M_\odot$ may be indicated.
We note that even higher bar masses ($M_{\rm BAR} \geq
5\times 10^{10} M_\odot$) make it difficult to achieve a flat
rotation curve interior to the solar radius and thus are not viable.

Turning now to the optical depth to the LMC (Fig.~2), we find as
expected that a thin disk makes a negligible contribution to $\tau_{\rm LMC}$,
while a thick disk can provide a significant contribution
(toward the LMC $\tau \propto h$).  In fact, a model with a heavy
bar and a thick disk can account for both the bulge and LMC rates
without recourse to MACHOs in the halo.  We also mention that
the LMC microlensing rate is small enough
that a significant part of it ($\sim 0.5\times 10^{-7}$)
could be due to microlensing by objects in the LMC itself \cite{lmclens}.

Our most striking result is that all viable models have
a significant halo component.  This can be traced to the flat-rotation curve
constraint (even though it is very conservative) and the upper
limit to local projected mass density \cite{1/r}, and is essentially
independent of the bulge optical depth.  That is,
while $\tau_{\rm BULGE}$ provides us with information about the disk
and bulge components of our galaxy, the halo parameters are relatively
insensitive to it.  This can be seen in Fig.~3,
where the local halo density in acceptable models is
around $5\times 10^{-25}\gcmm3$ with more or less the same
uncertainty as previous estimates (see e.g., Refs.~\cite{localhalo}).

Even more than the uncertainty in the observed LMC microlensing rate,
imprecise knowledge of the galactic model dominates the
uncertainties in determining the MACHO fraction of the halo.
However, in most viable models, an all-MACHO halo results
in an optical depth to the LMC that is many times that observed (Fig. 2),
and a significant nonbaryonic halo component seems indicated (Figs.~3 and 4).
The viable models with the largest MACHO fraction have thin disks,
small bar masses, large core radii and high local circular velocity.

To summarize:  (1)  A single component of the galaxy
cannot account for both the bulge and LMC events.
For example, a spherical halo predicts an LMC rate that is
slightly higher than the bulge rate; a thin disk cannot account for
either the bulge rate or the LMC rate.   A thick disk can
explain the LMC events, but not the bulge events.
(2)  The most promising model for explaining the high
bulge rate is an asymmetric bulge (bar) that lenses itself \cite{barlens}
with a lesser contribution from a thin disk.
(3) Viable models of the galaxy have a significant
halo component and the LMC rate expected for an all-MACHO
halo is many times that observed.  (4) While
the present data cannot preclude an all-MACHO halo, it appears
that the fraction of the halo in MACHOs is less than 30\%.
This is consistent with searches for faint halo stars
that indicate that their contribution
to the halo mass is small (provided that the mass function of halo
stars is ``smooth;'' see e.g., Ref.~\cite{bahcall}).

If the bulk of the halo is not in the form of MACHOs, what is it?
While it is not impossible that it could be baryonic, in
a more diffuse form, e.g., clouds of neutral gas \cite{jetzer},
cold dark matter is a more compelling possibility.  In an
$\Omega = 1$ cold dark matter model,  the naive expectation for
the baryonic mass fraction in our galactic halo is $f_B = \Omega_B$.
Based upon primordial nucleosynthesis $0.01h^{-2} \le\Omega_B
\le 0.02h^{-2}$ \cite{copi}, which implies $f_B \simeq 0.04-0.2$.
The expected baryon mass fraction for models with an
admixture of massive neutrinos ($\Omega_\nu \sim 0.2-0.3$)
is only slightly higher.  For cold dark matter models
with a cosmological constant, $\Omega_{\Lambda}\simeq 0.8$,
$f_B \simeq 0.1-0.2$.  If the baryonic halo has undergone
moderate dissipation, the baryonic mass the baryon
fraction in the inner part of the galaxy can be
increased, though it is still expected to compose less than half
of the local dark matter density \cite{gates}.    From Fig. 4,
it is clear that the baryonic mass fraction in our halo implied from
microlensing is consistent with any of these scenarios which
provides further motivation for the ongoing
experimental efforts to directly detect neutralinos and axions
in our own halo.

\section*{Acknowledgments}
We thank John Bahcall and Scott Tremaine for helpful conversations.
This work was supported in part by the DOE (at Chicago and
Fermilab) and the NASA (at Fermilab through grant NAG 5-2788).
GG was also supported in part by an NSF predoctoral fellowship.

\newpage

\section*{Figure Captions}
\bigskip

\noindent{\bf Figure 1:}  Optical depth to the bulge for bar masses
of 1, 2, and 3$\times 10^{10}M_\odot$ (bottom to top)
and thick (broken) and thin (solid) disks as a function
of the local projected mass density $\Sigma_{\rm VAR}$
for models that satisfy the kinematic
constraints ($r_0 = 8.5\kpc$ and $v_c = 220\kms$).

\medskip
\noindent{\bf Figure 2:}  Optical depth to the LMC from an all-MACHO
halo (upper lines) and thick (broken) and thin disks (solid)
for bar masses of 1, 2, and 3$\times 10^{10}M_\odot$ (right
to left) as a function of $\Sigma_{\rm VAR}$ for
models that satisfy the kinematic constraints ($r_0=8.5\kpc$ and
$v_c=220\kms$).

\medskip
\noindent{\bf Figure 3:}  Distribution of local cold dark
matter density in viable models for $\Sigma_{\rm VAR} = 5-15$, 25-35,
45-55, and $65-75M_\odot\,{\rm pc}^{-2}$.  Since the
halo MACHO fraction in most viable models is small, the
local cold dark matter density is approximately equal to the total
local halo density.

\medskip
\noindent{\bf Figure 4:} Distribution of halo MACHO mass fraction
in viable models for $\Sigma_{\rm VAR} = 5-15$, 25-35, 45-55,
and $65-75 M_\odot\,{\rm pc}^{-2}$.


\begin{thebibliography}  {std}

\bibitem{pac} B.~Paczynski, {\it Astrophys. J.} {\bf 304}, 1 (1986).

\bibitem{eros} E.~Aubourg et al., {\it Nature} {\bf 365}, 623 (1993).

\bibitem{ogle} A.~Udalski et al., {\it Astrophys. J.} {\bf L69}, 426 (1994);
{\it ibid}, in press (1994); {\it Acta Astron.} {\bf 43}, 289 (1993);
{\it ibid} {\bf 44}, 165 (1994); {\it ibid} {\bf 44}, 227 (1994).

\bibitem{macho} C.~Alcock et al., {\it Nature} {\bf 365}, 621 (1993).

\bibitem{griest} K.~Griest, {\it Astrophys. J.} {\bf 366}, 412 (1991).

\bibitem{bulpred}  K.~Griest et al., {\it Astrophys. J.} {\bf 372},
L79 (1991); M.~Kiraga and B.~Paczynski, {\it ibid} {\bf 430}, L101 (1994).

\bibitem{machobulge} C.~Alcock et al., astro-ph/9407009.

\bibitem{gates} E.~Gates and M.~S.~Turner, {\it Phys. Rev. Lett.}
{\bf 72}, 2520 (1994).

\bibitem{griestnew} C.~Alcock et al., astro-ph/9411019.

\bibitem{megomodels} J.~Frieman  and R.~Soccimarro, {\it Astrophys. J.}
{\bf L23}, 431 (1994).

\bibitem{machoproc} W.~Sutherland, {\it Nuc. Phys. B} (Proceedings
of Neutrino 94, Eilat, Israel), in press (1994).

\bibitem{Zarits} D.~Zaritsky and S.~D.~M.~White, {\it Astrophys. J.},
in press (1994).

\bibitem{binney}  J.~Binney and S.~Tremaine, {\it Galactic Dynamics}
(Princeton University Press, Princeton, 1987).

\bibitem{frenk} C.~S.~Frenk and S.~D.~M.~White, {\it Mon. Not. R. astr. Soc.}
{\bf 193}, 295 (1980).

\bibitem{expt} See e.g., J.R.~Primack, D.~Seckel, and B.~Sadoulet,
{\it Ann. Rev. Nucl. Part. Sci.} {\bf 38}, 751 (1988); P.F.~Smith
and D.~Lewin, {\it Phys. Repts.} {\bf 187}, 203 (1990).

\bibitem{glxmodels} J.~N.~Bahcall, M.~Schmidt, and R.M.~Soneira,
{\it Astrophys. J.} {\bf 265}, 730 (1983);
J.~A.~R.~Caldwell and J.~P.~Ostriker, {\it ibid} {\bf 251}, 61 (1981).

\bibitem{manybars} L.~Blitz, in {\it Back to the Galaxy} (AIP Conference
Proceedings 278), eds. S.~S.~Holt and F.~Verter, (AIP, New York, 1993), p.
98 (1993); J.~Binney, {\it ibid}, p.~87 (1993).

\bibitem{dwek} E.~Dwek et al., NASA/GSFC preprint (1994).

\bibitem{barmass}  S.~M.~Kent, {\it Astrophys. J.} {\bf 387}, 181 (1992).

\bibitem{GWK} G.~Gilmore,  R.~F.~G.~Wyse, and  K.~Kuijken, {\it
Ann. Rev. Astron. Astrophys.} {\bf 27}, 555 (1989).

\bibitem{lumdisk} See e.g., J.N.~Bahcall, {\it Astrophys. J.}
{\bf 276}, 169 (1984); K.~Kuijken and G.~Gilmore, {\it ibid}
{\bf 376}, L9 (1991).  Taking a smaller value for $\Sigma_{\rm LUM}$
would allow for larger $\Sigma_{\rm VAR}$ and hence
larger $\tau_{\rm BULGE}$; however, taking $\Sigma_{\rm LUM}
=0$ only increases $\tau_{\rm BULGE}$ by $0.3\times 10^{-6}$.

\bibitem{sigma0} J.N.~Bahcall, C.~Flynn, and A.~Gould, {\it Astrophys.
J.} {\bf 389}, 234 (1992); K.~Kuijken and G.~Gilmore, {\it ibid}
{\bf 367}, L9 (1991) and Refs.~\cite{glxmodels}.

\bibitem{1/r} Such a model (referred to as a Mestel
disk \cite{binney}) produces a flat rotation curve in the
plane of the galaxy; however, in order to
account for a rotation velocity of $220\kms$, a local surface mass
density of $220 M_\odot\,{\rm pc}^{-2}$ is required.
Gould [preprint OSU-TA-16/94 (1994)] has recently revisited these models,
because a thin Mestel disk ($h\sim 0.3\kpc$) whose local surface
density is about $220 M_\odot\,{\rm pc}^{-2}$ can:
(i) marginally account for the bulge rate; (ii) marginally account for
the LMC rate; and (iii) produce a flat rotation curve ($v = 220\kms$)
in the galactic plane without a halo.  However, such a model would not produce
a flat rotation curve outside the galactic plane nor would it explain
the warping of the disk and the flaring of neutral gas; moreover,
it is in severe conflict with kinematic studies that indicate
$\Sigma_{\rm TOT}(1\kpc )$ is at most $100M_\odot\,{\rm pc}^{-2}$
\cite{sigma0}.

\bibitem{lenses}  Our expression for $\tau$ when applied to the
bulge implicitly assumes that all the mass density in the
disk and bulge is available for lensing.  However, the bulge lenses
are certainly not bright stars.  Zhao et al. \cite{barlens}
find that correcting for this is a small effect.  To account
for the fact that the lens are unlikely to be bright stars
we simply do not include the contribution of the fixed luminous
disk to $\tau_{\rm BULGE}$.

\bibitem{barlens} B.~Paczynski et al., Princeton Observatory
preprint POP-573 (1994); H.~Zhao, D.~N.~Spergel and R.~M.~Rich,
astro-ph/9409022.

\bibitem{lmclens} K.~Sahu, {\it Nature} {\bf 370}, 275 (1994).

\bibitem{localhalo} Refs.~\cite{gates,glxmodels}; M.S.~Turner,
{\it Phys. Rev. D} {\bf 33}, 886 (1986); R.~Flores,
{\it Phys. Lett. B} {\bf 215}, 73 (1988).

\bibitem{bahcall} J.~Bahcall et al., {\it Astrophys. J.} {\bf 435},
L51 (1994).  In fact, A.~de Rujula et al. (astro-ph/9408099) have
raised the question of where the brown dwarf stars are.

\bibitem{jetzer} See e.g., F.~De Paolis et al., astro-ph/9411016;
astro-ph/9411018.

\bibitem{copi} C.~Copi, D.N.~Schramm, and M.S.~Turner, {\it Science},
in press (1995).

\end{thebibliography}
\end{document}